# Towards defining reference materials for extracellular vesicle size, concentration, refractive index and epitope abundance.

## Authors

Joshua A. Welsh[1]*, Edwin van der Pol[2,3,4], Britta A. Bettin[2,4], David R. F. Carter[5], An Hendrix[6], Metka Lenassi[7], Marc-André Langlois[8,9,10], Alicia Llorente[11], Arthur S. van de Nes[12], Rienk Nieuwland[2,4], Vera Tang[8,9], Lili Wang[13], Kenneth W. Witwer[14]†*, Jennifer C. Jones[1]†*

*Corresponding authors: JAW; joshua.welsh@nih.gov; KWW; kwitwer1@jhmi.edu, JCJ; jennifer.jones2@nih.gov
†Senior authors

## Affiliations

[1]Translational Nanobiology Section, Laboratory of Pathology, National Cancer Institute, National Institutes of Health, Bethesda, USA
[2]Vesicle Observation Center, Amsterdam UMC, location AMC, University of Amsterdam, Amsterdam, the Netherlands
[3]Biomedical Engineering & Physics, Amsterdam UMC, University of Amsterdam, Amsterdam, the Netherlands
[4]Laboratory of Experimental Clinical Chemistry, Department of Clinical Chemistry, Amsterdam UMC, location AMC, University of Amsterdam, Amsterdam, the Netherlands
[5]Department of Biological and Medical Sciences, Faculty of Health and Life Sciences, Oxford Brookes University, Oxford, UK
[6]Laboratory of Experimental Cancer Research, Department of Human Structure and Repair, Ghent University, Ghent, Belgium; Cancer Research Institute Ghent, Ghent, Belgium
[7]Institute of Biochemistry, Faculty of Medicine, University of Ljubljana, Ljubljana, Slovenia
[8]University of Ottawa Flow Cytometry and Virometry Core Facility, Ottawa, Canada
[9]Department of Biochemistry, Microbiology and Immunology, Faculty of Medicine, University of Ottawa, Ottawa, Canada
[10]Ottawa Center for Infection, Immunity and Inflammation, Ottawa, Canada
[11]Department of Molecular Cell Biology, Institute for Cancer Research, Oslo University Hospital, Norway, Research, Department of Molecular Cell Biology, Oslo, Norway
[12]VSL Dutch Metrology Institute, Delft, the Netherlands
[13]National Institute of Standards and Technology (NIST), 100 Bureau Drive, Stop 8312, Gaithersburg, MD, 20899, USA
[14]Departments of Molecular and Comparative Pathobiology and Neurology, The Johns Hopkins University School of Medicine, MD, USA

## Funding

JAW and JCJ are supported by the U.S. National Institutes of Health, National Cancer Institute, 1ZIA-BC011502, and the Intramural Research Program of the National Institutes of Health (NIH), National Cancer Institute, and Center for Cancer Research (JAW). JCJ acknowledges NIH ZIA BC011502, NIH ZIA BC011503, NIH U01 HL126497, NIH R01 CA218500, NIH UG3 TR002881, and the Prostate Cancer Foundation. KWW acknowledges NIH DA040385, DA047807, AI144997, MH118164, and UG3CA241694, as well as the Michael J. Fox Foundation. JAW is an ISAC Marylou Ingram Scholar 2019-2023. VAT is an ISAC Shared Resource Lab Emerging Leader 2018-2022. ML is supported by the Slovenian Research Agency (ARRS) projects P1-0170 and J3-9255. ALL is supported by The Norwegian Research Council and by the South-Eastern Norway Regional Health Authority. EvdP is supported by the Netherlands Organisation for Scientific Research—

Domain Applied and Engineering Sciences (NWO-TTW), research program VENI 15924. BB is supported by European Metrology Research Programme Joint Research Project 18HLT01: METVES II.

**Conflict of Interest**

JAW, JCJ, KWW, ML, RN, ALL, AH report no conflict of interest. MAL is CEO of ViroFlow Technologies, Inc. EvdP is co-founder, CSO and shareholder of Exometry BV.

**Acknowledgements**

The authors thank John P. Nolan and George Daaboul from Cellarcus Biosciences and NanoView, respectively, for providing reagent information and constructive feedback.

**Abstract**

Accurate characterization of extracellular vesicles (EVs) is critical to explore their diagnostic and therapeutic applications. As the EV research field has developed, so too have the techniques used to characterize them. The development of reference materials is required for the standardization of these techniques. This work, initiated from the ISEV 2017 Biomarker Workshop in Birmingham, UK, and with further discussion during the ISEV 2019 Standardization Workshop in Ghent, Belgium, sets out to elucidate which reference materials are required and which are currently available to standardize commonly used analysis platforms for characterizing EV size, concentration, refractive index, and epitope expression. Due to their predominant use, a particular focus is placed on the optical methods nanoparticle tracking analysis and flow cytometry.

**Introduction**

The connection of EVs to many aspects of human health and disease surged a global interest in the development of EV-based biomarkers and therapeutics [1]. The use of EVs requires techniques which are able to reliably characterize size distribution, concentration, and epitope expression, amongst others. EVs can be studied by ensemble and single particle techniques. An ensemble technique measures one or more properties of a bulk EV population e.g. enzyme-linked immunosorbent assay (ELISA), Western blot, total protein/lipid/nucleic acid concentration, or bead-based flow cytometry. Single particle techniques characterize EVs one-by-one. Examples of single particle techniques are flow cytometry, nanoparticle tracking analysis (NTA, a commercialized name of single particle tracking (SPT)), electron microscopy (EM) and resistive pulse sensing (RPS). Ensemble techniques are often scalable, sensitive, and therefore compliant with routine clinical applications. For example, the first widely used screening test for HIV was based on ELISA [2]. Despite these benefits, the use of ensemble techniques has only limited abilities to improve understanding of disease-specific EV subsets due to the component diversity of many body fluids. These often contain many different particles with overlapping size and physical properties, such as lipoprotein particles, protein complexes, small platelets and EVs from many cell types. When applied to putative disease-specific EV subsets, therefore, interpretation of ensemble techniques may rely heavily on the purity of the EV preparation. For example, a Western blot positive for an EV marker may be informative for a highly purified EV population, but not for a neat biological fluid, where the signal could come from soluble protein. In contrast, single particle techniques have the potential to identify single EVs and differentiate between EV subsets (EVs with common molecular profile or cargo) and other non-EV particles. If the technique is high-throughput and allows sufficiently multiplexed phenotypic characterization, it could even obviate the need for separation, a particularly important consideration for clinical applications..

The detection of single EVs in body fluids is challenging because EVs (1) are heterogeneous in size, with the majority having a diameter <200 nm; (2) are also heterogeneous in composition, biogenesis and origin; (3) have a low (<1.42) refractive index (RI); and (4) often co-exist with non-EV components that overlap in biochemical composition and/or physical properties [3]. The diameter distribution of EVs in normal human plasma and urine have been shown to range between 50 nm to >1,000 nm [4, 5]. Because instruments detecting single EVs, such as flow cytometers, differ in sensitivity, and because only a fraction of the EVs exceeds the detection threshold, minute differences in the lower limit of detection will strongly affect the measured concentration of EVs [5]. Understanding the performance strengths and limitations of single EV characterization techniques is crucial to generate reliable and reproducible data and can also help to identify approaches to improve these analysis techniques and assays [6].

There is a growing awareness that the reliability and reproducibility of EV measurements in general needs to improve. To become clinically relevant, EV measurements must be standardized. To this end, reference materials and reference procedures require development. Rigor and standardization efforts in the field are represented by publication of "minimal requirements" (MISEV2014, MISEV2018) and improved recording of pre-analytics in "EV-TRACK" [6-9], and also by standardization studies on EV concentration measurements by tunable RPS [10], flow cytometry [11, 12], NTA [13], and functional coagulation assays [14]. Recently, the ISEV-ISAC-ISTH EV flow cytometry working group published a 'minimum information for the reporting of an EV flow cytometry experiment' (MIFlowCyt-EV) standard reporting framework, in an effort to increase the transparency and reproducibility of flow cytometry experiment protocols and reporting [15].

While the need for reference materials is becoming increasingly recognized, the nomenclature and purpose of reference materials within the EV field is currently poorly defined, with some nomenclature commonly misused. Here, we focus on understanding what is meant by a reference material, what types of reference material are required by the EV field, and how reference materials and EVs in general should ideally be characterized and reported. The majority of single EV measurements are currently performed using the optical characterization methods NTA and flow cytometry [16]. We will therefore focus on standardization of these analysis methods, and draw several comparisons with non-optical measurement techniques.

**Assessing common EV characterization techniques**

A range of analysis techniques have been used to characterize EVs. **Table 1** provides a comparison of popular EV analysis techniques, indicating their ability to provide measurements of diameter, immunophenotyping data, concentration, refractive index, single particle detection, detect all individual EVs, have a derivable sensitivity limit, and achieve a large sampling of particles (>10,000 events). As one of the first characteristics of an EV analysis technique is assessing whether it is detecting single or multiple (bulk) particles, techniques are split into these two respective cateogories.

***Diameter distribution determination*** Single particle methods are needed to generate accurate size distributions. For this reason, NTA, flow cytometry, and electron microscopy have been widely used for EV diameter distribution reporting [16]. Newer methods such as resistive pulse sensing, super-resolution microscopy, and interferometric reflectance imaging sensing (IRIS) are also beginning to be utilized [17, 18]. Bulk methods such as DLS that can be prone bias due to not measuring single particles are used increasingly less frequently.

***Molecular phenotyping.*** Bulk analysis techniques, such as Western blots, ELISAs, and bead capture assays, have been widely used and indeed instrumental in the field to date, associating EV phenotype (molecular cargo) with function. However, bulk analysis methods cannot convey if a particular analyte is in or on all EVs or just a subset, reveal the distribution of markers within a positive subset, or identify the size distribution or concentration of the positive subset. They therefore lack the ability to characterize the heterogeneity of the EV population, which could be seen as critical for some of their intended uses in clinical chemistry. For this reason, there is a strong impetus to develop single particle analysis techniques, and an increasing number of platforms have become available. As seen in **Table 1**, only electron microscopy and super-resolution microscopy are capable of phenotyping single EVs of the smallest diameter. These are specialized and low-throughput (time-intensive) methods, though, that can analyze only a small portion of the population, possibly neglecting low abundance particles such as large EVs. High-throughput methods are therefore desirable for single particle phenotyping. NTA can technically be used for high-throughput fluorescence-based phenotyping, but low detection sensitivity and fluorophore bleaching have limited its application. Flow cytometry is another high-throughput possibility, but a lack of minimum procedural and reporting guidelines for single EV flow cytometry, combined with variable equipment sensitivities, settings, and staining methodologies, has led to a general lack of reproducible data. This has only recently been address in the form of the MIFlowCyt-EV framework [19].

***Concentration determination:*** The determination of concentration is a multifactorial measurement as it quantifies the amount of signal within a set volume. How well the measurement signal is being differentiated from background e.g. are all of the EVs detectable, and the ability to accurately determine the analysed volume both play a role in accurate concentration determination. If a technique is unable to detect all particles within a population, the correct interpretation of a particle concentration measurement therefore requires quantitation of the limit of detection of the measurement device. This then allows the reporting of concentration within a set sensitivity window i.e. $3 \times 10^7$ particles mL$^{-1}$, limit of detection = 157 nm, that can be reproduced.

The lower limit of detection (limit of sensitivity), is the threshold at which signal (such as light scatter for NTA) from a particle of given size can no longer be discriminated from the background. The limit of sensitivity can be described in numerous ways, with some easier to quantitate than others. In the above example with NTA, the limit of sensitivity could be reported in terms of the number of photons needed to resolve a signal from background, or as the diameter of a particle with a given refractive index. Despite this, it is also one of the few single-particle measurement techniques that also does not have a definable sensitivity limit with regard to light scatter intensity or fluorescence intensity in standard units.

***Sensitivity versus resolution.*** Distinguishing between 'sensitivity' and 'resolution' is important. While 'sensitivity' describes the ability to detect a signal, 'resolution' describes the ability to distinguish one signal from another. **Figure 1**, depicts hypothetical results when three techniques are used to detect particles with diameters of 75, 100, and 125 nm: measured either as single populations (left column, **Figure 1A, C, E**) or when mixed together (right column, **Figure 1B, D, F**) to approximate a heterogeneous mixture of EVs. In this example, each particle population contributes an equal number of events. Although each technique is sufficiently *sensitive* to detect the individual populations, *resolution* differs substantially. The method in **Figure 1A**, **B** has high resolution. **Figure 1C**, **D** shows a low-resolution method. The technique in **Figure 1E**, **F** loses resolution as particles (and their signal) become smaller. This latter pattern is typical of detection methods such as flow cytometry, NTA, and RPS. Clearly, our perception of a population size distribution can be skewed if resolution is not taken into account

***Current techniques: an overview.*** Each currently available platform for single EV characterization has strengths and weaknesses. NTA is widely used and can provide size, concentration, and even phenotyping data (when combined with affinity-linked fluorescence). However, it is unable to detect all EVs, and there is currently no published protocol to determine a lower limit of sensitivity in standard units for fluorescence or irrespective of refractive index for size. This is currently also true for the combination of IRIS (interferometry) and fluorescence. Resistive pulse sensing (RPS) is also incapable of detecting all individual EVs, as the pores used for the measurement have discrete size ranges. However, RPS can measure tens of thousands of events in a short period of time, and its measurement sensitivity can be deduced using currently available traceable size standards. Since RPS does not allow affinity-based phenotyping, EV size and concentration can be interpreted correctly only for populations well separated from potential co-isolates, such as lipoproteins in plasma. This is also the case for other techniques where affinity-based phenotyping is not or cannot be done. Currently, only single-EV flow cytometry combines abilities for sizing, concentration measurement, affinity-based phenotyping, and high throughput with calibration into standard units to provide a limit of sensitivity for each parameter. While not all techniques have a discernable limit of sensitivity that can be derived from a calibration and expressed in standard units, it is possible to perform indirect assessments of sensitivity using validation materials. It is also possible to track performance using quality controls. These are discussed in detail below.

**Background to reference materials**

Metrology is the science of measurement, fundamental to the definition of reference materials as it is the scientific study of measurements. Metrology is regulated by the International Bureau of Weights and Measurements (BIPM) and ensures international unification of physical measurements. The metrology field is established internationally with regards to nomenclature and regulatory agencies. The BIPM operates under the exclusive supervision of the International Committee for Weights and Measurements (CIPM), which established the International System of (SI) Units in 1960. The SI unit, known as the metric system, is the international measurement standard. The SI unit is recognized in nearly 50 countries, with the CIPM disseminating and modifying the definition of SI units as technology progresses. The International Organization for Standards (ISO) is an independent, non-government organization that interacts with BIPM. ISO is the largest developer of international standards and provides common standards between over 160 countries. Over 22,910 standards have been published to date. An example of an ISO standard is the definition of accuracy, ISO 5725-1:1994, 'Accuracy (trueness and precision) of measurement methods and results.' National metrology institutes, such as the US National Institutes for Standards and Technology (NIST), develop certified reference materials which are traceable to the SI unit using ISO standards. Measurement traceability is the property of a measurement result whereby the result can be related to a reference through a documented unbroken chain of calibrations, each contributing to the measurement uncertainty.

The SI unit is made up of seven base units that define 22 derived units with special names and symbols, **Figure 2A**. The use measurements of certified reference materials within laboratories can be traced back to the SI units, **Figure 2B**. An example is a gold size standard being made and characterized in comparison to an international size standard, **Figure 2C**. This gold size standard is used by manufacturers of polystyrene size standards which are characterized in relation to the gold size standard by electron microscopy. These polystyrene size standards can then be bought commercially and used to calibrate laboratory instrumentation. The working methods and reference materials with the laboratory instrumentation are then used to characterize assays output e.g. the size of an EV. As reference materials continue to be cross-calibrated, their accuracy and stability decreases, **Figure 2B**. This is due to every measurement containing a certain amount of uncertainty and over increasing time and variable conditions stability decreases, **Figure 3**.

'Reference material' is a generic term that refers to any material that is sufficiently homogeneous and stable with respect to one or more properties, and which has been established to be fit for its intended use in a measurement process [20]. A 'certified reference material' is characterized by a metrologically valid procedure for one or more of its properties and is accompanied by a certificate providing the values of the specified property, a statement of metrological traceability and associated uncertainty. Metrologically valid procedures for production and certification are outlines in ISO Guide 34 and 35, certificate contents ISO Guide 31. Reference materials can have different applications e.g. calibration, validation, quality control, etc. Each of these applications can require varying degrees of accuracy in their reference material characterization e.g. international standards, certified reference materials, working reference materials, and more.

For a reference material to be traceable back to an SI unit, the uncertainty of a measurement must be known, **Figure 3**. The uncertainty of a measurement is the quantification of doubt about the measurement and is based upon the standard deviation and bias of a measurement. The standard deviation describes the precision of a measurement due to random error. The measurement bias describes the trueness of a measurement, which can be affected by systematic error. 'Error' in both systematic and random error describe the difference between the measured value and the 'true value' of the property being measured. The trueness and precision of a measurement, defined by ISO 5725, equate to the accuracy of a measurement, **Figure 4**. A measurement whereby the result is low in systematic error but high in random error is considered high in trueness but lacking in precision. A measurement whereby the result is high in systematic error but low in random error is considered low in trueness but high in precision. Understanding these concepts is critical for the development of EV reference materials, assessing EV analysis equipment, and general investigation and characterization of EVs.

**Reference material categories for standardization of EV characterization techniques**

The development and utilization of reference materials may fall into one of the following categories: calibration, validation, and quality control. The need for certified vs uncertified reference materials for these needs will vary, as discussed below.

*Calibration* is the conversion of an arbitrary unit scale to a scale in standard units. For example, converting the fluorescence intensity scale of a flow cytometer from arbitrary units to a scale of fluorophore number. Calibration requires well characterized reference particles. Ideally, the calibration reference materials would also be certified, so as to give a traceable measurement back to SI units and limiting the potential for bias in calibration accuracy. The development of certified calibration reference particles is crucial for instrument measurement calibration, instrument sensitivity quantification, and in turn instrumentation standardization and comparisons. Calibration reference material allow standardized reporting and consequently, validation of published EV studies between assays, instruments, and laboratories. The development of accurate calibration reference materials also allows the development of standardized quality control reference materials.

Calibration reference materials do not necessarily need to mimic all characteristics of the downstream analysis particles, such as EVs. Materials such as polystyrene beads, hollow organosilica beads, liposomes, etc, could all be conceivably used as size standards, so long as the measurement technique sizing ability is based on physical size and not on other properties such as light scatter intensity (which would also depend on refractive index). Calibration reference materials for some parameters, such as fluorescence and light scatter, typically require a set of standards for multiple populations. Other parameters, such as concentration, can be based on a single calibration reference material.

*Validation* assesses an assay's sensitivity and specificity by using a known sample, such as a positive or negative control. For example, a scale could be calibrated to diameter units using a 100 nm certified diameter standard. To validate the calibration of the scale, a 150 nm certified diameter standard could be analyzed to check that the population appeared at 150 nm on the scale. An assay's specificity for detecting CD41a-positive population could be validated by using a population of CD41a certified positive particles for a positive result. Inversely, the assay specificity for negative results could be validated using CD41a certified negative particles for a negative result. Validation reference materials appear to be the most sought-after within the EV field in order to test detection methods and assays. These types of reference materials can be used as positive and negative controls within assays by having previously characterized properties, such as protein expression, concentration, diameter distribution, composition, etc. While in some cases calibration reference materials can also be used for assay validation, validation should not be carried out using the same reference material as used for calibration. Despite great potential for the use of EV reference materials for assay validation, currently only uncertified EV reference materials are commercially available. Uncertified reference materials can lack accuracy in characterization measurements, and depending on how they're used for their downstream application can lead to bias in sample characterization e.g. using an incorrectly assigned size standard to calibrate a RPS device will lead to measurement inaccuracy of downstream sample diameter measurements. Many commercially available reference materials have not been rigorously quantified or otherwise characterized, and sometimes not at all. These types of reference materials are therefore not standard, and may not be traceable across different detection methods, and thus result in variable data across instrumentation. Their use as quality control reagents or assay development reagents on a single platform may, however, be useful.

The development of certified EV reference materials is currently impeded due to limitations in sensitivity and resolution of commercially available analysis techniques. Most current commercially available techniques are not able to detect and/or phenotype the full range of the EV population as single particles or and/or cannot provide a large enough sampling of the population to produce robust measurements. Furthermore, many of the techniques that can provide a higher sampling of the EV population, as seen in **Table 1**, cannot yet be calibrated to provide a quantitative sensitivity limit or traceable measurement. The measurement bias of these techniques can therefore not be accounted for, and the "product specification" metrics provided along with reference materials will likely be inaccurate.

***Quality control*** assesses the performance of an assay ('repeatability' and 'reproducibility') to verify that it produces consistent results over time. Repeatability assesses whether a measurement is consistent when performed: at the same location; using the same measurement procedure; by the same observer; using the same measuring instrument, used under the same conditions; and repeated over a short period of time. 'Reproducibility' assesses whether measurements are consistent when conducted by different individuals, at different locations, and with different instruments. Quality control can be assessed intra-assay with replicate samples or inter-assay with calibration or validation reference materials. Quality control for all measurement techniques is essential and is best quantified in standard units e.g. "the detection sensitivity decreased from 100 nm to 150 nm."

**Commercially available EV reference materials for standardization of EV analysis techniques**
A number of efforts are being undertaken to develop reference materials as EV mimetics for use as validation of assays and quality controls. Some come in the form of synthetic materials meant to mimic biological materials, such as hollow organosilica beads and liposomes, while others are derived from biological sources [21] [22-24].

Synthetic materials such as hollow beads have the benefit of utilizing characteristics of EVs with fine control, such as low refractive indices and a core shell structure, whilst being in the format of a tightly defined population which will likely be unambiguous in its detectability using optical methods. Synthetic standards such as these are likely much easier to develop into certified reference materials due to being stable, homogeneous populations that are more amenable to analysis and assigning traceable metrics. However, hollow silica beads have not yet been developed to display or contain proteins or other molecules that would be useful as positive controls for assays and testing of reagents.

Biological reference materials that have been proposed in the literature include virus particles, cell culture-derived (including engineered) EVs, bacterial outer membrane vesicles, and biofluid-derived EVs (from urine, plasma, serum, etc.). Each of these biological reference material types has strengths and limitations that are application dependent. The generation of biological reference materials to a standard that allows certification is challenged by the large number of parameters involved and the lack of sensitive instrumentation that can provide traceable measurements. Perhaps the most achievable goal is that these EV reference materials be developed and reported in calibrated units that state the detection window of the reported metrics. Efforts to produce high-quality biological reference materials with calibrated measurements are recent, and have been demonstrated in the form of retroviruses and recombinant EVs [23, 24].

The development of assays, reagents, and detection methods strongly requires that biological reference materials are commercially available for assay validation. The lack of commercial availability is in part due to the lack of calibration and reporting standards within the field. The currently available reference materials that could be used for assay validation are shown in **Table 2**. Recently, an inter-societal flow cytometry working group with members from ISEV, ISAC, and ISTH made an effort to overcome the lack of experimental and reporting standards for flow cytometry. This effort resulted in the MIFlowCyt-EV reporting framework, which was published as a position paper in the *Journal of Extracellular Vesicles,* [19]. The majority of the MIFlowCyt-EV framework is applicable to most optical analysis techniques, despite being developed for flow cytometry. Utilization of the framework would result in a large step forward for the field, not only in standardizing reporting and being able to validate findings, but also in starting to characterize and create standard commercially available biological reference materials.

A comparison of commercially available EV reference materials and the level to which they are characterized is collated into **Table 2**. The criteria from this table are metrics that we would propose as a bare minimum when considering the use of these materials for downstream characterization using an optical characterization method and reporting the results. These available reference materials include EVs, EV-sized retroviruses, and liposomes. An absolute minimum of reporting EV studies is describing the method used to separate the EVs [9].

At this time, no commercial sources of biological reference materials in **Table 2** provide proof of purity or indication of purification method. These reagents may, therefore, contain soluble proteins or free nucleic acids, that could have effects on downstream assays if used without further purification. The reporting of size distribution is a common and important factor when using reference materials for the characterization or validation of an assay using an optical analysis technique. While most studies report a diameter statistic, such as mean or median, few provide the variance of this diameter distribution, and no manufacturer currently provides the limit of sensitivity for the technique quantifying the diameter. As seen in **Table 2**, despite limitations in sensitivity quantification and standardization (**Table 1**), many of the diameter distributions of currently available biological EV mimetics utilise NTA. An approximate RI is reported for three reference materials, two of which are viral particles. Buoyant density was provided by only one manufacturer. Two manufactures provide known surface proteins and their brightness in calibrated units, while another with an extensive range of EV types gives no indication of known surface markers. Critically, no commercially available reference materials intended for use as EV references provide the limit of sensitivity for their measurement metrics.

**Development of new reference materials for standardization of frequently studied EV characteristics**

*Refractive index determination*

The refractive index of a material determines how much light is scattered from it. The refractive index has no effect on measurements from non-optical techniques such as RPS. For optical techniques that detect light scatter (e.g. flow cytometry, NTA, DLS, NanoView), the refractive index strongly influences the particle measurements in the detectability of the particle or derivation of the particle diameter. The refractive index is therefore an important metric to be provided with a reference material if it is intended for use in optical analysis techniques. Current literature suggests that while the RI of EVs is lower than reference materials such as silica, it is variable, **Figure 5**. Indeed, the effective refractive index of vesicles will never be a single number due to a vesicle being a core-shell model, where the ratio of the shell (membrane) to the core (cytosolic portion) increases as EV size decreases. Smaller EVs will therefore have larger effective RIs than larger EVs, such that RI cannot be reported with a single metric.

Calibration of optical analysis equipment for refractive index approximation generally requires utilizing particle scatter physics in the form of Mie theory. While refractometers exist, they tend to be designed for liquids or films and are therefore not optimized for single particles or polydisperse mixtures of spheres. The application of Mie theory modelling requires calibration with spherical particles of known diameter and refractive index in order to fit predicted scattering models to detection equipment [25-27]. Many certified polystyrene reference particles exist. These are monodisperse and have low variance and a reported refractive index. Unfortunately, few alternative materials are as well characterized. For some materials, such as silica, refractive index is assumed based on published literature from measurements of silica films. These materials may thus be variable in composition quality, with few certified reference materials available. The refractive indices currently provided with certified reference particles tend to be reported to two decimal places and with no indication of variance; they are therefore not a traceable metric.

Assay validation for sensitivity to detect particles of low refractive index remains difficult. Analyses of polystyrene particles were initially used to validate the sensitivity of optical analysis instrumentation. However, it was soon recognized that polystyrene, having a much higher refractive index, scatters far more light than most biological particles, and therefore is not an appropriate EV mimetic. Since then, focus has been directed to the use of silica particles, which, while lower in RI than polystyrene, still scatter more light than EVs. Liposomes have more recently been suggested as a better material for validating equipment sensitivity due to the similarity in RI to EVs; however, it is difficult to produce monodisperse liposomes with low variance at small diameters. Liposomes also tend to be multi-laminar if created using filters, thus increasing refractive index and not being representative of EVs with a single phospholipid bilayer.

The ability to calibrate an instrument's axis for refractive index will however continue to require the use of light scatter modelling and not be a traceable measurement until certified traceable refractive index reference particles exist. Without the ability to compare or validate refractive index measurements to a certified reference material, the accuracy of reported measurements will remain unknown.

### *Epitope measurement*

Phenotyping, defined for our purposes here as classifying EVs by presence of surface markers, is a powerful tool in determining cell derivation and function of EVs. Methods to phenotype EVs vary from qualitative (e.g. Western blots), to semi-quantitative (e.g. ELISA, bead-capture assays, fluorescent NTA, EM), to quantitative (e.g. super-resolution microscopy, flow cytometry). Reference materials that display a known quantity of a particular epitope (i.e., target of an affinity reagent such as an antibody, peptide, or aptamer) will therefore be important for a wide variety of assays and analysis platforms.

*Crucial: limit of sensitivity.* As for other types of measurements, it is critical to know the limit of sensitivity of phenotyping assays. To achieve a quantitative assay, this limit should be reported in standard units so that (for example) amount of bound fluorescent antibody can be expressed in molecules of equivalent soluble fluorophore. Reporting data in quantitative units (e.g. molecules of equivalent soluble fluorophore) instead of as relative expression or in arbitrary units (fluorescence intensity) allows for comparison of data across scientific institutions and platforms. Where this is impossible, many assays are still able to indicate the relative amount of an epitope, comparing single EVs or EV samples. In cases for which an epitope is undetectable, a lack of detection should not be confused with proof that an epitope is truly absent from an EV. Instead, the result should be reported in context of the limit of sensitivity, along with other relevant variables of the particular assay.

***Fluorescence calibration.*** Development of reference materials for calibration of epitope abundance is contingent on the analysis technique. Currently, optical techniques are most widely used, relying predominantly on fluorescence for the quantification of phenotype, so fluorescence calibrators are required in the form of molecules of equivalent soluble fluorophore. These currently exist only in the form of uncertified reference beads in the range of 3-7 $\mu$m in diameter that lack traceability. The size and fluorescence of these beads reflects their originally intended use in cellular analyses and requires extrapolation of their values to be applicable to the majority of EVs (i.e. 10,000 MESF to <100 MESF). While the true fluorescence value (i.e. accuracy) of these brighter reference materials has not be validated for very dim particles, initial small particle calibration studies using these beads suggest that they are precise and produce concordant data irrespective of instrumentation [28-30]. This is entirely expected, as fluorescence scales linearly with the number of molecules present, and existing measurements can assess if a fluorescence detector scales linearly. The use of fluorescence calibration is highly recommended by the ISEV-ISAC-ISTH working group for flow cytometry, as outlined in the MIFlowCyt-EV framework [19, 28].

## *Size distribution*

The size (diameter, radius) of EVs is one of their defining characteristics. Current evidence suggests that EVs from some sources have a log-normal diameter distribution, **Figure 6** [5]. That is, abundance scales inversely with size up to the peak of the distribution at a small size, which is itself determined by biophysical characteristics of lipid bilayer-delimited particles. Currently, no high-throughput analysis method is capable of sizing the full range of EVs (**Table 1,** [5]). A large limitation of the current literature is that reported EV diameters or diameter distributions do not state the limit of sensitivity of the detection equipment. For example, it is often stated that EV populations have a mean diameter of 100 nm. This is a distorted perception of the true EV size distribution, as in many cases the majority of EVs (small EVs) are undetectable and thus excluded from the distribution. A single metric to describe size distribution of EV samples—such as mean, mode, median, or percentile—is insufficient since it is biased by the sensitivity limit of the instrumentation, **Figure 6**. Techniques that rely on the composition of EVs (i.e., refractive index) for detection and inference of diameter, such as flow cytometry, NTA, DLS, and interferometry, will be biased by the refractive index of individual particles. Illuminate wavelength also contributes to sensitivity of optical techniques. An NTA instrument with a 405 nm laser will produce a slightly different size distribution compared with an instrument with a 640 nm laser due to the sensitivity of the instrument differing in light scatter from the particles at the two wavelengths. For these reasons, optical methods require calibration in order to determine their limit of sensitivity.

***Flow cytometry: light scatter and fluorescence.*** The approximation of particle diameter from light scatter or fluorescence intensity in flow cytometry requires the use of light scatter calibration and fluorescence calibration, respectively [25, 27, 31]. Certified size standards are available for various sizes and compositions. These certified size standards are homogeneous and traceable in their diameter measurement, made from synthetic materials providing stability, and are often reported with a density and refractive index. Flow cytometer light scatter calibration can utilize certified diameter standards for the approximation of particle diameter. This measurement uses both the diameter of the certified reference particle and the non-traceable refractive index measurement. The accuracy of diameter approximated using flow cytometer light scatter is therefore multi-factorial. Fluorescence calibration for EV size inference has been demonstrated with a liposome reference material of known diameter that is labelled with the same intercalating membrane dye as a population of EVs [31]. The fluorescence diameter measurement is also multifactorial and relies upon the accuracy of the liposome diameter, the detectability of the liposome population, and the fluorescence intensity of the liposome population as compared with the EV population. Importantly, both sizing methods allow for quantifying the limit of sensitivity in a standard unit of measurement irrespective of the instrument, allowing platform-independent comparisons.

***NTA: considerations and limitations.*** NTA sizes particles not by a single intensity measurement, as in flow cytometry, but rather by tracking the Brownian motion of particles over time (multiple measurements). Size is then inferred from the Stokes-Einstein equation. NTA does, however, rely on optical intensity to track particles over a sufficient length of time to derive an accurate size. Determining the limit of sensitivity for NTA would therefore require light scatter modelling or fluorescence calibration, depending on tracking mode, as well as some way to account for 1) the movement of particles in and out of the field of view, 2) changing intensities, and 3) the ability of the instrument to track them. In light scatter mode, intensity depends on refractive index and illumination wavelength. The enumeration of particles is then affected by the camera's varying noise, fluctuating at a pixel level over time, over which the software must identify and track a particle over several time frames. In fluorescence mode, additional information includes the amount of dye and rate of photobleaching. Currently, there is no demonstrated method to express limit of sensitivity for NTA in standard units, irrespective of factors such as refractive index of fluorescence intensity, such that comparisons could be made across instrumentation.

In summary, certified reference materials for size measurements by platforms like flow cytometry and NTA do not actually rely, or solely rely, on the physical size of the reference material. Certified reference materials that cater to calibrated fluorescence measurements and refractive index are therefore required. Currently, certified sizing standards are most applicable to calibration of non-optical methods, such as RPS, allowing for statements about accuracy and limit of sensitivity. RPS is thus a useful orthogonal technique to assess optical sizing. Similarly, cryo-electron microscopy and similar techniques, despite being low- throughput, can be used to assess EV morphology across the full range of sizes and correlate findings between detection methods.

The development of standard EV reference materials that can be used across optical techniques for validation of size distributions using light scatter requires knowledge of refractive index, which is not fixed across EV diameter and is not currently a traceable measurement using available methods. Similar standard reference materials for fluorescence requires knowledge of fluorescence intensity in standard units, which, while feasible, is not yet a traceable measurement and is currently only compatible with flow cytometers that are mostly unable to detect the whole EV population. A potential method of attempting to standardize optical size measurements is under investigation. A European standardization study (METVES II) attempts to make assay validators using hollow organosilica spheres of known sizethat mimic the core-shell structure of EVs, **Table 2**. This project aims to help standardize the field by using EV light scatter and fluorescence mimetics to validate instrument sensitivity.

### *Concentration measurement*

Particle concentration may be a useful parameter to normalize EV input into an *in vitro* or *in vivo* assay, or even in the diagnostic setting. Despite being reported in almost every EV publication, EV concentration is one of the most difficult metrics to derive due to the systematic and random errors that are involved. No high-throughput analysis method is capable of detecting all EVs with a single configuration. Importantly, the limit of sensitivity for techniques such as flow cytometry, NTA, and RPS is generally not reported, even though this cutoff is crucial to knowing how many of the smallest and most abundant EVs are detected. The reported concentration of EVs is therefore likely one of the least accurate metrics in the literature. While certified reference materials for concentration measurements already exist, an accurate concentration measurement of a sample requires detection of all particles. This is not the case for techniques such as RPS, NTA, and flow cytometry. The solution to this problem is, however, relatively simple: reporting detectable events within a given detection window, using a calibrated instrument. If an instrument has been calibrated, a concentration can be reported within a given detection window e.g. "$1.3 \times 10^9$ particles per mL were detected between 80-300 nm."

Calibrating EV analysis equipment to determine what can and cannot be detected in standard units is easier on some instruments than others. Instruments using RPS can be calibrated with size standards irrespective of optical characteristics such as refractive index. Optical techniques such as flow cytometry and NTA require light scatter calibration and/or fluorescence calibration to define their limits of sensitivity and their detection window. Both flow cytometry and NTA have a number of variables to account for, and well-defined, and ideally certified reference materials would be used for their calibration. These limitations are the same as those discussed previously for epitope measurement and size distribution.

The development of EV reference materials with a known concentration is, therefore, heavily dependent on the instrumentation being used to quantify concentration and whether those instruments are capable of detecting the whole population. When they are not capable of detecting the whole population, the concentration should be reported within a defined size range of the EV population. Given that current high-throughput methods are unable to detect the full EV population, it is unlikely that an accurate concentration measurement for the full EV population can be reported at this stage for current commercially available reference materials. If the limits of sensitivity are reported with reference materials, the concentration measurement can still, however, be normalized across instrumentation.

**Discussion**

The characterization of reliable EV reference materials requires the calibration of measurements obtained from any EV analysis technique. Calibration is also critical for the characterization of samples reported in published data. The utilization of current reference materials and development of new reference materials relies upon several factors. These include (a) continuing efforts in the EV field to develop educational resources and workshops for understanding and teaching the need and utilization of standardization procedures, (b) journals enforcing minimal criteria for reporting experiments using EV analysis techniques, and (c) encouraging industry to utilize and develop robust reference materials for calibration and quality control of EVs.

The development of platform-independent reference materials is needed to facilitate cross-platform standardization. While flow cytometry and NTA were a particular focus of this piece, emerging optical techniques such as super-resolution microscopy and interferometry are also becoming more widely used. The standardization of these analysis techniques will be aided by the development of reference materials for flow cytometry and NTA, since each of these techniques measures optical signals.

While the EV field currently and severely lacks standardization, this problem is recognized by the field. An example of an initiative aimed at development of traceable reference materials is METVES II, described in Text Box I. The International Society for Extracellular Vesicles (ISEV) in 2019 initiated a 'Rigor and Standardization' subcommittee to coordinate task forces across several pertinent areas, one of which is 'EV Reference Materials'. Recently, an ISEV workshop, hosted in Ghent, Belgium was dedicated to Rigor and Standardization. Initiatives to standardize reporting have also been made in the form of MISEV, EV-TRACK, and MIFlowCyt-EV. Progress towards standardization is thus gaining momentum via concerted efforts.

**Textbox** "METVES II (https://www.metves.eu), is a European metrology project in which metrology institutes, companies, and academia collaborate to achieve standardization of concentration measurements of EVs in clinical samples, such as plasma and urine [32]. METVES II focuses on standardization of EV concentration measurements by flow cytometry, since it is already available in hospitals and can characterize single EVs at high throughput (thousands/sec). All aspects of flow cytometry, including flow rate, fluorescence, and light scatter

(size, refractive index) need to be calibrated to produce reliable and reproducible results. Toward this goal, dedicated and traceable reference materials are being developed that combine physical properties resembling those of EVs. These materials will include stable particles with (i) diameters between 50 nm and 1,000 nm, (ii) concentrations in the range of $10^9$ - $10^{12}$ particles mL$^{-1}$, (iii) a visible fluorescence intensity between 100 and 100,000 molecules of fluorophore, and (iv) a refractive index (RI) in the range of 1.37 – 1.42. Three types of reference materials will be developed: hollow organosilica beads (HOBs) [33], monodisperse liposomes, and low-RI solid particles. The size and concentration of the developed reference particles will be traceably characterized in SI units [34, 35]. It is hoped that this project will deliver a single reference material to calibrate all relevant properties involved in EV flow cytometry measurements."

**References**

1. Tkach, M. and C. Thery, *Communication by Extracellular Vesicles: Where We Are and Where We Need to Go.* Cell, 2016. **164**(6): p. 1226-1232.
2. Alexander, T.S., *Human Immunodeficiency Virus Diagnostic Testing: 30 Years of Evolution.* Clin Vaccine Immunol, 2016. **23**(4): p. 249-53.
3. Coumans, F.A.W., et al., *Methodological Guidelines to Study Extracellular Vesicles.* Circ Res, 2017. **120**(10): p. 1632-1648.
4. Arraud, N., et al., *Extracellular vesicles from blood plasma: determination of their morphology, size, phenotype and concentration.* J Thromb Haemost, 2014. **12**(5): p. 614-27.
5. van der Pol, E., et al., *Particle size distribution of exosomes and microvesicles determined by transmission electron microscopy, flow cytometry, nanoparticle tracking analysis, and resistive pulse sensing.* J Thromb Haemost, 2014. **12**(7): p. 1182-92.
6. Consortium, E.-T., et al., *EV-TRACK: transparent reporting and centralizing knowledge in extracellular vesicle research.* Nat Methods, 2017. **14**(3): p. 228-232.
7. Cvjetkovic, A., J. Lotvall, and C. Lasser, *The influence of rotor type and centrifugation time on the yield and purity of extracellular vesicles.* J Extracell Vesicles, 2014. **3**.
8. Clayton, A., et al., *Summary of the ISEV workshop on extracellular vesicles as disease biomarkers, held in Birmingham, UK, during December 2017.* Journal of Extracellular Vesicles, 2018. **7**(1).
9. Thery, C., et al., *Minimal information for studies of extracellular vesicles 2018 (MISEV2018): a position statement of the International Society for Extracellular Vesicles and update of the MISEV2014 guidelines.* Journal of Extracellular Vesicles, 2018. **7**(1).
10. Vogel, R., et al., *A standardized method to determine the concentration of extracellular vesicles using tunable resistive pulse sensing.* J Extracell Vesicles, 2016. **5**: p. 31242.
11. Lacroix, R., et al., *Standardization of pre-analytical variables in plasma microparticle determination: results of the International Society on Thrombosis and Haemostasis SSC Collaborative workshop.* J Thromb Haemost, 2013.
12. Cointe, S., et al., *Standardization of microparticle enumeration across different flow cytometry platforms: results of a multicenter collaborative workshop.* J Thromb Haemost, 2017. **15**(1): p. 187-193.
13. Vestad, B., et al., *Size and concentration analyses of extracellular vesicles by nanoparticle tracking analysis: a variation study.* J Extracell Vesicles, 2017. **6**(1): p. 1344087.
14. Hisada, Y., et al., *Measurement of microparticle tissue factor activity in clinical samples: A summary of two tissue factor-dependent FXa generation assays.* Thromb Res, 2016. **139**: p. 90-7.
15. Welsh, J.A., et al., *MIFlowCyt-EV: a framework for standardized reporting of extracellular vesicle flow cytometry experiments.* Journal of Extracellular Vesicles, 2020. **9**(1): p. 1713526.
16. Gardiner, C., et al., *Techniques used for the isolation and characterization of extracellular vesicles: results of a worldwide survey.* J Extracell Vesicles, 2016. **5**: p. 32945.
17. Daaboul, G.G., et al., *Digital Detection of Exosomes by Interferometric Imaging.* Scientific Reports, 2016. **6**(1): p. 37246.
18. Lennon, K.M., et al., *Single molecule characterization of individual extracellular vesicles from pancreatic cancer.* J Extracell Vesicles, 2019. **8**(1): p. 1685634.
19. Welsh, J.A., et al., *MIFlowCyt-EV: a framework for standardized reporting of extracellular vesicle flow cytometry experiments.* Journal of Extracellular Vesicles, 2020.
20. May W, P.R., Beck II C, Fassett J, Greenberg R, Guenther F, Kramer G, Wise S, Gills T, Colbert J, Gettings R, MacDonald B, *Definitions of Terms and Modes Used at NIST for Value-Assignment of Reference Materials for Chemical Measurements ; NIST Special Publication 260-136*. 2000, U.S. Government Printing Office: Washington, DC
21. Varga, Z., et al., *Hollow organosilica beads as reference particles for optical detection of extracellular vesicles.* J Thromb Haemost, 2018.
22. Valkonen, S., et al., *Biological reference materials for extracellular vesicle studies.* Eur J Pharm Sci, 2017. **98**: p. 4-16.
23. Tang, V.A., et al., *Engineered Retroviruses as Fluorescent Biological Reference Particles for Nanoscale Flow Cytometry.* bioRxiv, 2019: p. 614461.

24. Geeurickx, E., et al., *The generation and use of recombinant extracellular vesicles as biological reference material.* Nature Communications, 2019. **10**(1): p. 3288.
25. van der Pol, E., et al., *Single vs. swarm detection of microparticles and exosomes by flow cytometry.* J Thromb Haemost, 2012. **10**(5): p. 919-30.
26. van der Pol, E., et al., *Refractive index determination of nanoparticles in suspension using nanoparticle tracking analysis.* Nano Lett, 2014. **14**(11): p. 6195-201.
27. Tang, V.A., et al., *Engineered Retroviruses as Fluorescent Biological Reference Particles for Small Particle Flow Cytometry.* bioRxiv, 2019: p. 614461.
28. Welsh, J.A. and V.A. Tang, *Light scatter and fluorescence calibration allow standard comparisons of small particle data between different instruments.* bioRxiv, 2019: p. 796961.
29. Morales-Kastresana, A., et al., *Labeling Extracellular Vesicles for Nanoscale Flow Cytometry.* Sci Rep, 2017. **7**(1): p. 1878.
30. Morales-Kastresana, A., et al., *High-fidelity detection and sorting of nanoscale vesicles in viral disease and cancer.* Journal of Extracellular Vesicles, 2019. **8**(1).
31. Stoner, S.A., et al., *High sensitivity flow cytometry of membrane vesicles.* Cytometry A, 2016. **89**(2): p. 196-206.
32. Nicolet, A., et al., *Inter-laboratory comparison on the size and stability of monodisperse and bimodal synthetic reference particles for standardization of extracellular vesicle measurements.* Measurement Science and Technology, 2016. **27**(3).
33. Varga, Z., et al., *Hollow organosilica beads as reference particles for optical detection of extracellular vesicles.* Journal of Thrombosis and Haemostasis, 2018. **16**(8): p. 1646-1655.
34. Meli, F., et al., *Traceable size determination of nanoparticles, a comparison among European metrology institutes.* Measurement Science and Technology, 2012. **23**(12): p. 125005.
35. Varga, Z., et al., *Towards traceable size determination of extracellular vesicles.* Journal of extracellular vesicles, 2014. **3**: p. 10.3402/jev.v3.23298.
36. Sultanova, N., S. Kasarova, and I. Nikolov, *Dispersion Properties of Optical Polymers.* Acta Physica Polonica A, 2009. **116**(4): p. 585-587.
37. Daimon, M. and A. Masumura, *Measurement of the refractive index of distilled water from the near-infrared region to the ultraviolet region.* Appl Opt, 2007. **46**(18): p. 3811-20.
38. Malitson, I.H., *Interspecimen Comparison of the Refractive Index of Fused Silica*,†.* Journal of the Optical Society of America, 1965. **55**(10): p. 1205-1208.
39. Gardiner, C., et al., *Measurement of refractive index by nanoparticle tracking analysis reveals heterogeneity in extracellular vesicles.* J Extracell Vesicles, 2014. **3**: p. 25361.
40. van der Pol, E., et al., *Absolute sizing and label-free identification of extracellular vesicles by flow cytometry.* Nanomedicine, 2018. **14**(3): p. 801-810.
41. de Rond, L., et al., *Refractive index to evaluate staining specificity of extracellular vesicles by flow cytometry.* Journal of Extracellular Vesicles, 2019. **8**(1).

**Table 1 – Comparison of highly reported EV characterization methods.** For diameter, immunophenotyping, concentration, and refractive index ticks depict whether the instrument provides or has been demonstrated to prove a particular measurement metric. For diameter, immunophenotyping, concentration, and refractive index, crosses indicate the instrument does not, is not able to, or has not been demonstrated in published literature to provide a particular element at the time of writing this review.

**Table 2 – Comparison of basic optical parameter characterization of commercially available reference materials.** Information was collated using manufacturer websites and product sheets that were openly available. It is possible further is known about these products but that information is not freely/openly available at the time of writing this review.

**Figure 1 – The importance of resolution.** A) demonstrated the detection of particles with a consistently high resolution ($\mu$ = 75, 100, 125 nm, $\sigma$ = 3, 3, 3), B) shows the cumulative diameter distribution of particles from plot A. C) demonstrated the detection of particles with a consistently low resolution ($\mu$ = 75, 100, 125 nm, $\sigma$ = 15, 15, 15), D) shows the cumulative diameter distribution of particles from plot C. E) demonstrated the detection of particles with a typical detection technique resolution, whereby it decreases as the signal becomes smaller ($\mu$ = 75, 100, 125 nm, $\sigma$ = 15, 10, 3), F) shows the cumulative diameter distribution of particles from plot E. All populations have 10,000 particles.

**Figure 2 – Traceability to the SI unit.** A) shows the seven base units of the SI unit, the outer circle shows the base unit measurement, the middle circle shows the measurement unit, the inner circle shows the measurement unit symbol. B) Hierarchy of traceability from the working methods and reference materials to the SI unit. C) shows an example of an EV measurement using RPS back to the SI unit.

**Figure 3 – Parameters involved in characterizing a certified reference material.**

**Figure 4 – Visualizing the difference between trueness and precision.**

**Figure 5 – Dispersion properties of polystyrene, silica and water from wavelengths of 400-800 nm.** Dispersion properties of polystyrene, silica, and water were calculated using the Sellmeier equations for published materials [36-38]. Median refractive index (geometric mean in case of Gardiner *et al*) measurements for different EV sources were acquired from the literature [24, 26, 39-41].

**Figure 6 – Limitations of statistical metrics on partially resolved populations.** Shown is a hypothetical dataset with log-normal diameter distribution. Three limits of sensitivity (100 nm, blue; 150 nm, green; 300 nm,

red) are shown. The summary statistcs for events above these limits of sensitivity are shown in the corresponding colors in the right of the plot.

1
2

# Table 1

| | Technique | Can output diameter measurement? | Capable of phenotyping? | Can output concentration measurement | Can output refractive index measurement? | Measurements based on single particles | Capable of detecting full EV population | Can output quantitative sensitivity limit | >10,000 single particles per measurement |
|---|---|---|---|---|---|---|---|---|---|
| **Single Particle Detection Assay** | Nanoparticle tracking | ✓ | ✓ | ✓ | ✓ | ✓ | X | X | X |
| | Flow cytometry (single EV) | ✓ | ✓ | ✓ | ✓ | ✓ | X | ✓ | ✓ |
| | Electron microscopy | ✓ | ✓ | ✓ | X | ✓ | ✓ | ✓ | X |
| | Resistive pulse sensing (coulter) | ✓ | X | ✓ | X | ✓ | X | ✓ | X |
| | NanoView | ✓ | ✓ | ✓ | X | ✓ | X | X | X |
| | Super resolution microscopy | ✓ | ✓ | ✓ | X | ✓ | ✓ | ✓ | X |
| **Bulk Detection Assay** | Flow cytometry (bead-assay) | X | ✓ | X | X | X | X | ✓ | X |
| | Dynamic light scattering | ✓ | X | X | X | X | X | X | X |
| | Refractometer | X | X | X | ✓ | X | X | ✓ | X |
| | ELISA | X | ✓ | X | X | X | X | ✓ | X |
| | Western Blot | X | ✓ | X | X | X | X | ✓ | X |

# Table 2

| Company | ViroFlow Technologies Inc | ViroFlow Technologies Inc | Cellarcus Inc | Cellarcus Inc | Cellarcus Inc | Excytex | Excytex | HansaBioMed | Exosomics | Exometry |
|---|---|---|---|---|---|---|---|---|---|---|
| Reference material | Murine Leukemia Virus | Murine Leukemia Virus-sfGFP | Platelet-derived EVs | RBC-derived EVs | Liposomes (LIPO100) | Liposome (XCX-MIMIC) | Liposome (XCX-TARGET) | Cell-derived EVs | Cell-derived EVs | Hollow organosilica beads |
| Cell derivation | NIH3T3, DF-1, HEK293T | NIH3T3, DF-1, HEK293T | Human platelets | Human erythrocytes | NA | NA | NA | Variety of cell lines | (source not reported) | NA |
| Purity/purification method | | | | | | | | | | NA |
| Diameter range* | ✓ | ✓ | ✓ | ✓ | ✓ | | | | | ✓ |
| Vendor Analysis methods | Flow cytometry, NTA | Flow cytometry, NTA | Flow Cytometry, NTA, RPS | Flow Cytometry, NTA, RPS | Flow Cytometry, NTA, RPS | | | NTA | NTA | TEM, SAXS, Flow cytometry |
| Concentration measurement method* | Flow cytometry | Flow cytometry | Flow cytometry | Flow cytometry | Flow cytometry | | | Inferred from protein content (100 µg = 1x10$^{10}$) | NTA | |
| Refractive index (RI)* | ✓ | ✓ | | | | | | | | ✓ |
| Buoyant density (g/ml) | ✓ | ✓ | | | | | | | | Neutral in water |
| Lipid membrane | Cell membrane derived | Cell membrane derived | Cell membrane derived | Cell membrane derived | Synthetic | Synthetic | Synthetic | Cell membrane derived | Cell membrane-derived | NA |
| Intracellular cargo | ✓ | ✓ | | | | | | | | NA |
| Surface marker measurement* | ✓ | ✓ | ✓ | ✓ | ✓ | ✓ | ✓ | | | NA |
| Expression in calibrated units | ✓ | ✓ | ✓ | ✓ | ✓ | | | | | NA |
| Stated sensitivity limit (*=applicable metrics) | | | | | | | | | | |

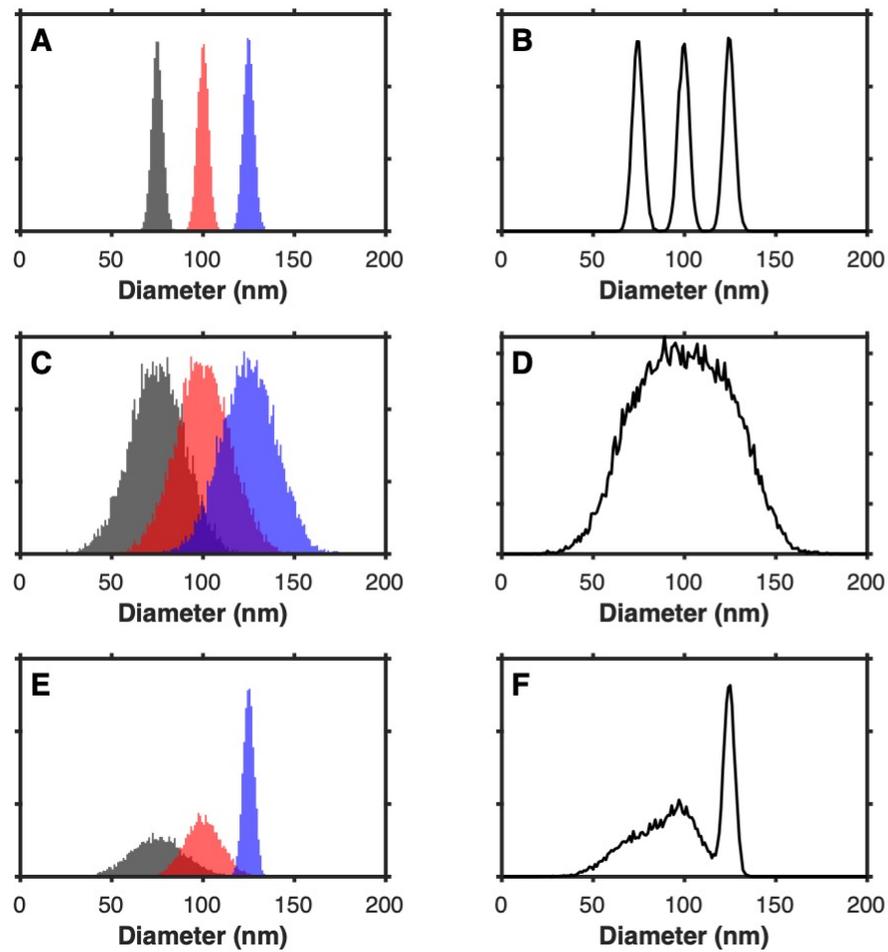

**Figure 1**

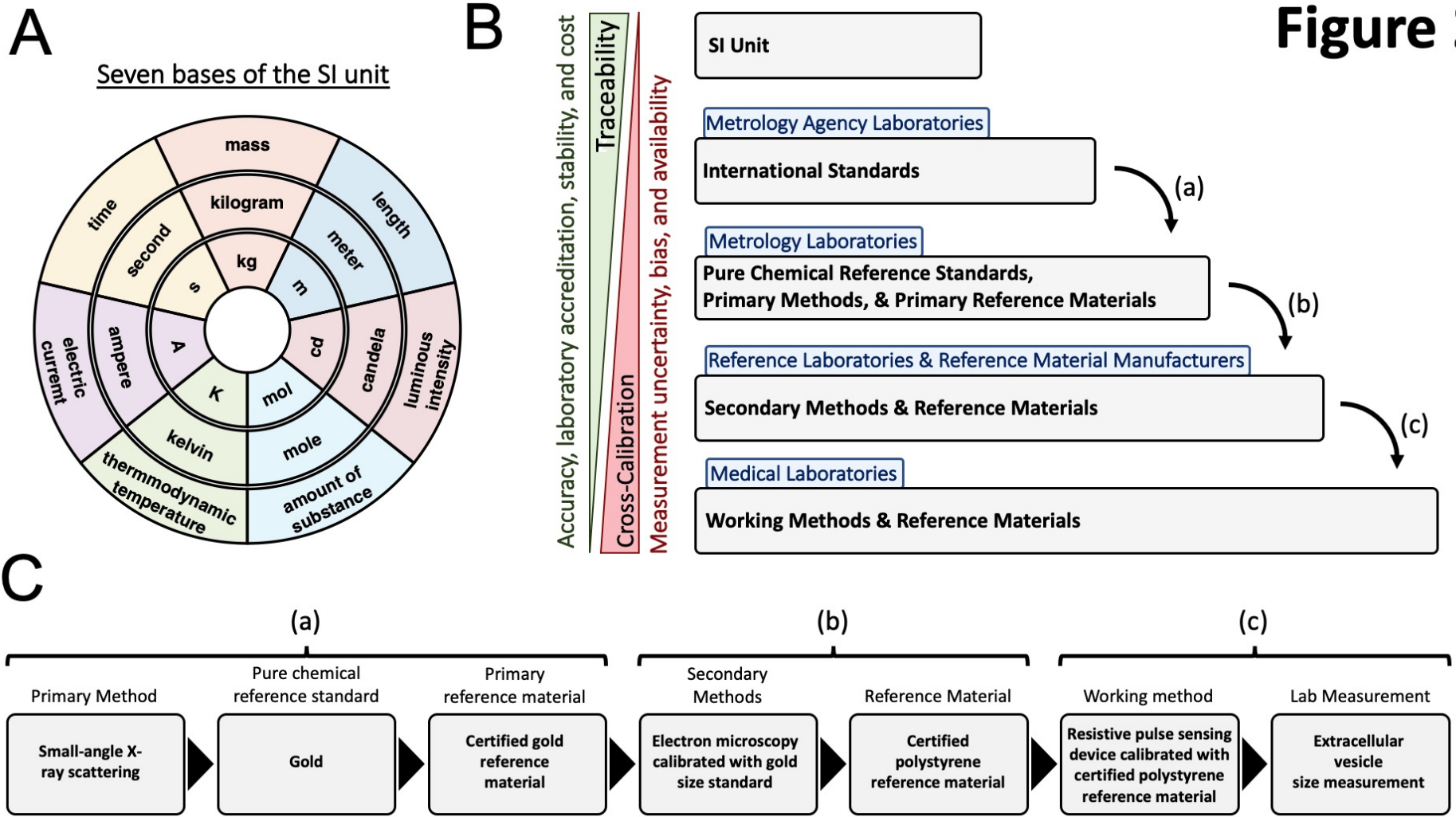

Figure 2

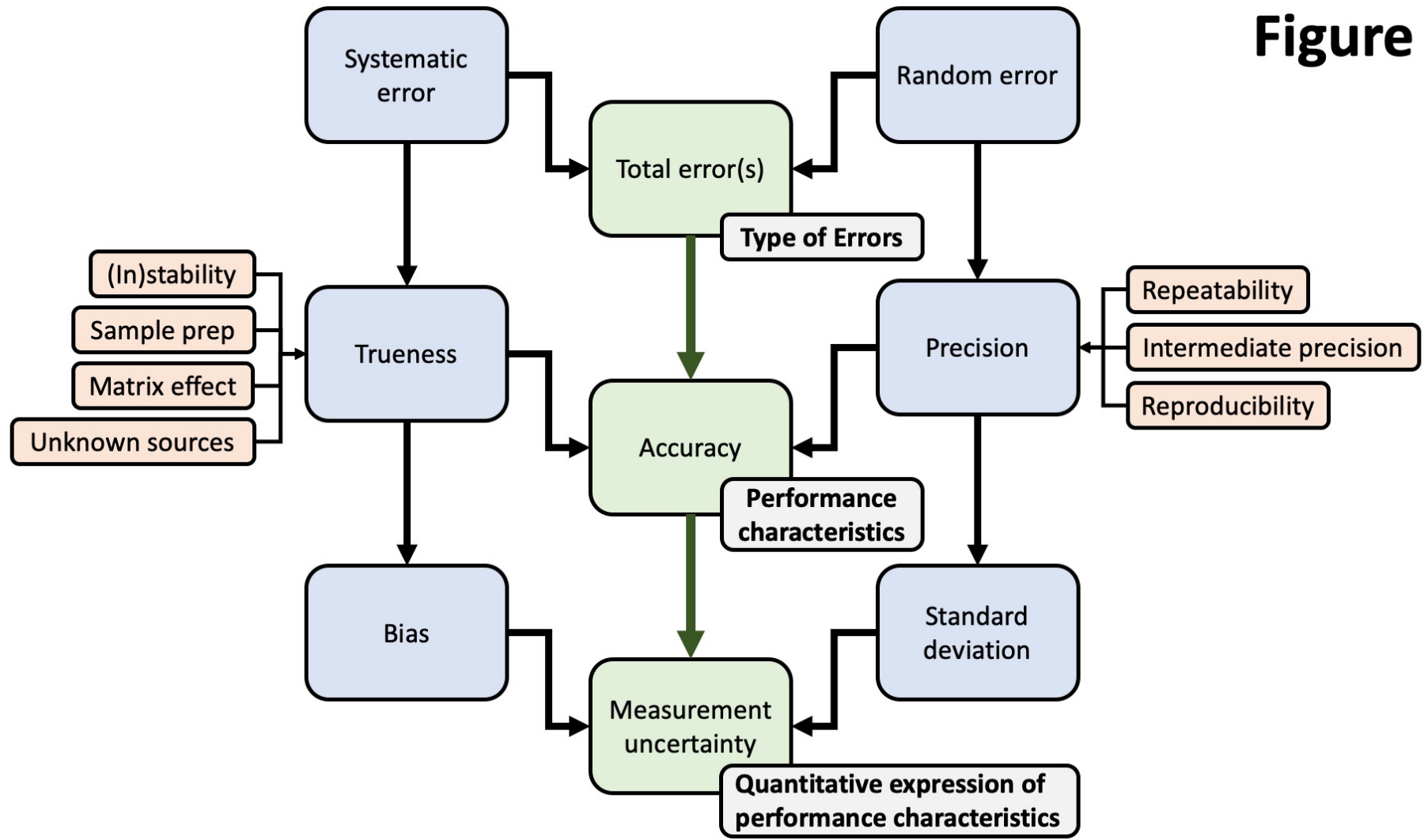

**Figure 3**

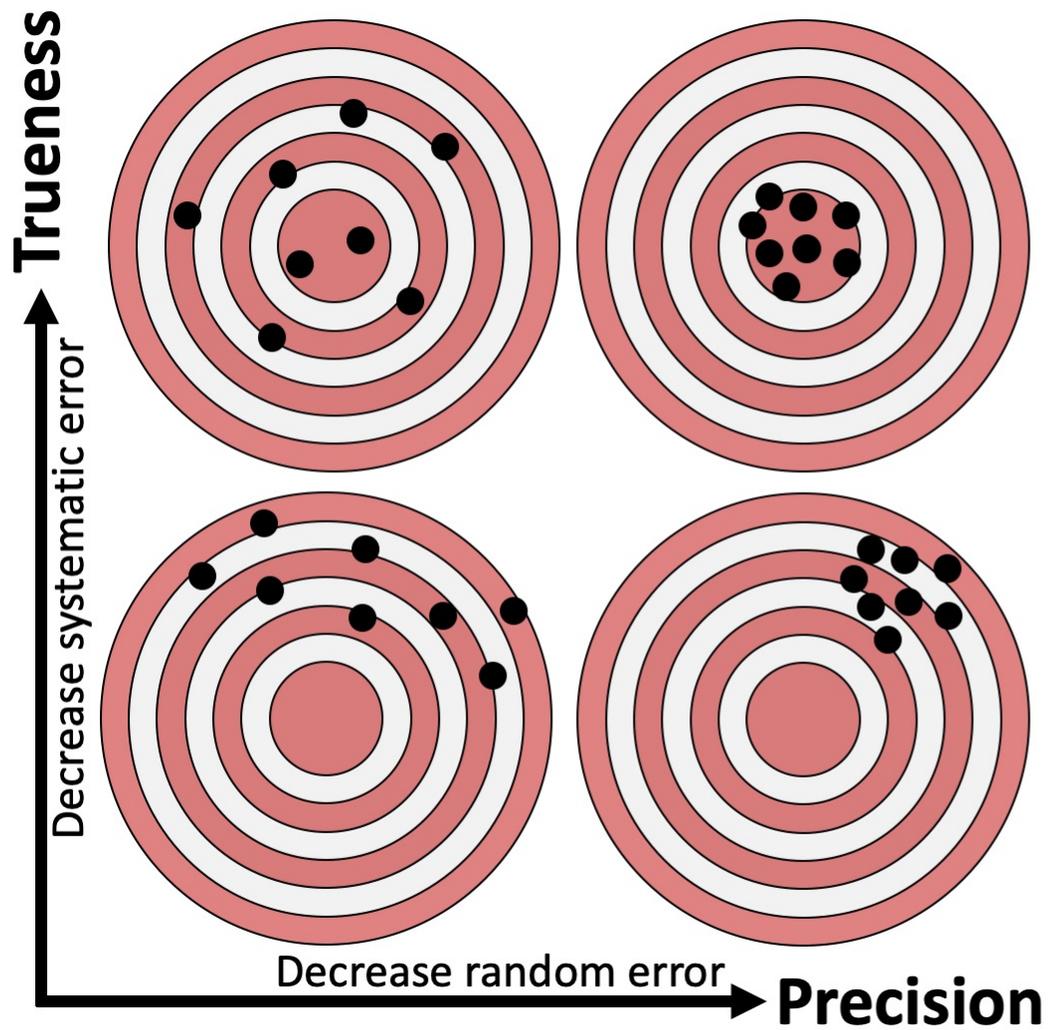

**Figure 4**

**Figure 5**

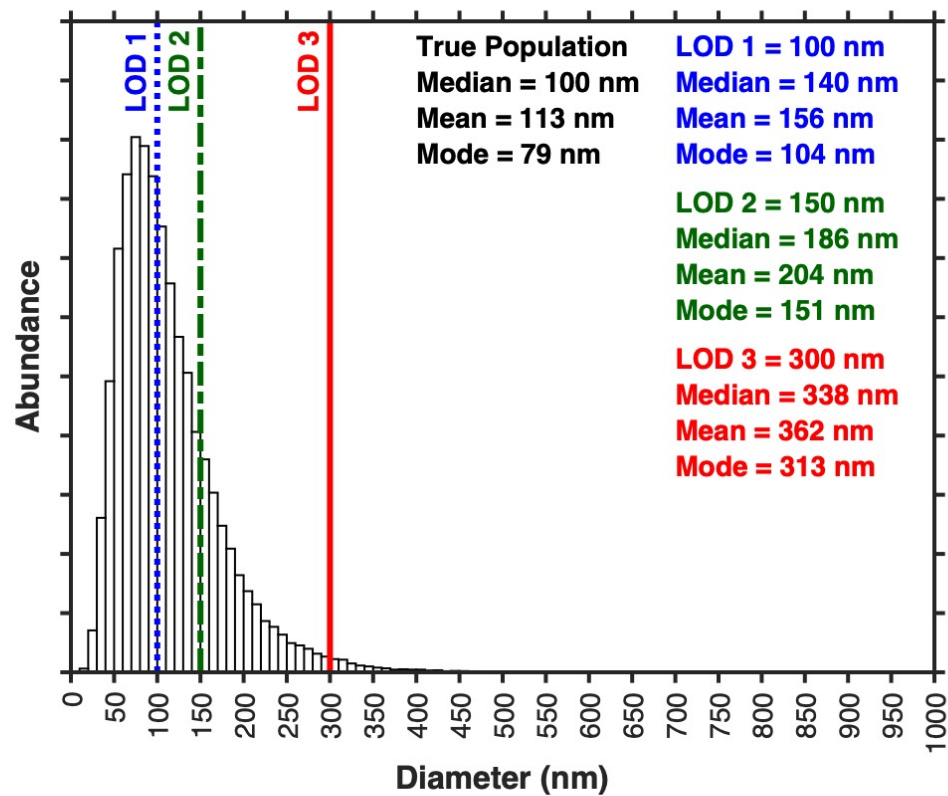

**Figure 6**